\documentclass[12pt]{article}
\usepackage{epsfig}

\catcode`\@=11
\catcode`\@=12

\title{{\small \hfill IMSc-2003/05/09}\\
\textbf{ Mesons: Relativistic Bound States with String Tension}}

\author{ Ramesh Anishetty\footnote{ramesha@imsc.res.in}~ and ~Santosh Kumar Kudtarkar\footnote{sant@imsc.res.in}\\
 The Institute of Mathematical Sciences, Chennai 600 113, India\\}

\date{\today}
\begin{document}
\maketitle

\begin{abstract}
\indent
 A systematic method of analysing Bethe-Salpeter equation using spectral representation
for the relativistic  bound state wave function is given. This has been explicitly applied in the
context of perturbative QCD with string tension in the $1 \over N$ expansion. We show that
there are only a few stable bound state mesons due to the small "threshold mass"(constituent mass) of quarks.
The asymptotic properties of the bound states are analytically analysed. The spectrum is derived
analytically and compared phenomenologically. Chiral symmetry breaking and PCAC results are
demonstrated. We make a simple minded observation to determine
the size of the bound states as a function of the energy of the boundstate.
\end{abstract}

%\noindent PACS numbers: 14.80.Hv, 11.15.-q, 11.15.Tk\\
\newpage

\section{Introduction}
 We address relativistic bound states which are due to a causal interaction kernel. 
Investigation of these systems is essential  to 
understand approximate
Goldstones such as the physical pion. Wick-Cutkosky(WC) model\cite{wick,cut} was one such model which 
was investigated in great detail, wherein they have presented a fairly general series expansion technique. 
Here we simplify and make their formalism 
more transparent and infact
we find that ``deeply bound states'', those whose binding energy is 
comparable or more than
the rest mass energy due to a very strong interaction kernel can be understood in a simpler way.
An important observation has been that the most general ``spectral'' representation \cite{ida} for the bound state
wave function exists and is very simple to work with. WC wave functions are a special class
of this representation.
  
  In this work we will study quark-antiquark($\bar q q$) bound states in the Bethe-Salpeter(BS) 
formalism in the context of the field theoretic model ($\sigma QCD$) proposed in \cite{ram}. 
 To recapitulate the 
essential points of this model, string tension term($\frac{\sigma}{k^4}$) was explicitly 
incorporated in perturbative 
QCD using auxillary fields such that ultraviolet renormalisation is assured. 
The ultraviolet(UV) behaviour remains the same as in QCD. The 
string tension($\sigma$) vanishes asymptotically in the UV limit. In this model we will be 
working in the leading $\frac{1}{N}$ approximation and  $g^2N$ is assumed to be small for all 
energies where $g$ is the QCD gauge coupling constant. The infrared singular confining part of the interaction is given by the string tension 
term. Our analysis is done in Minkowski(+,-,-,-) space.

 In $\sigma QCD$ with our approximations we have seen that the quark propagator has no pole 
and it does not have a simple pole structure\cite{sd} unlike in WC model. Consequently the BS equation
which involves the quark propagator has more algebraic complications. Even then the bound
state spectral representation is still valid and this enables us to perform analytic
calculations. Qualitatively we see that quark propagator poles are missing but they have
``threshold masses''\cite{sd}  which determines the onset of the imaginary  part of the propagator.
This is a more precise notion in our model corresponding to constituent mass of strong
interaction phenomenology.  
 For completeness we have presented the angular decomposition of the wave function in detail.
For brevity we have looked at single quark flavour system. Our analysis can equally well handle
cases of more than one flavour.

  In the BS bound state description of mesons we show that even in the presence of string tension
there are only a few number of stable mesons and this is a consequence of the existence of the the threshold
mass. There however are many unstable(complex energy) bound states and we have not made any attempt to 
study them systematically.

 Heavy quark bound systems under certain standard assumptions do reduce to non-relativistic 
Schrodinger theory bound systems. This is alluded to briefly as it is well understood in the 
literature. As for light mesons we derive the relationship between the mass of the pion and
the current quark masses consistent with PCAC. 

\section{Bethe-Salpeter equation}
\begin{figure}[htbp]
\begin{center}
\epsfig{file=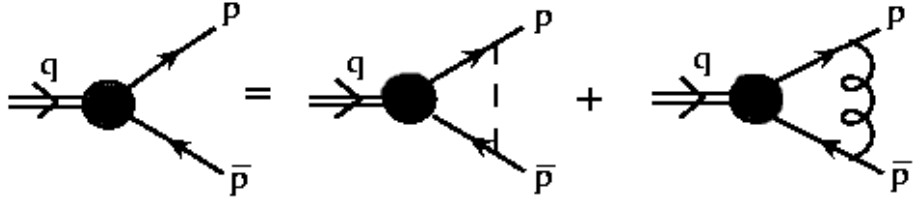,width=14cm,angle=0}
\end{center}
\caption{Bethe-Salpeter equation summing up ladder graphs.}
\label{bethfig}
\end{figure}

We address the quark-antiquark bound state problem in perturbative QCD
with string tension. As discussed in \cite{ram,sd} there are three
parameters in the theory, $\sigma N$,  $g^2 N$ and $1/N$ of which we
will treat  $\sigma N$ as a non-perturbative parameter and the latter
two perturbatively. The BS equation (Fig[\ref{bethfig}]) for the quark-antiquark
bound state in the $1/N$ expansion sums only the  ladder graphs
of '$\sigma$ exchange' (eq.(\ref{BS0})) where  quark antiquark
propagators are the non-perturbative propagators obtained by summing
the rainbow  '$\sigma$ exchange' \cite{sd}.

\begin{equation}
\phi(p,\bar p)= S(\bar p)\gamma_{\mu} \int \frac{d^4 k}{i (2\pi)^4}(
 \frac{\sigma N  \phi((p-k),(\bar p+k))}{ (p-k)^4}-  
\frac{ g^2 N \phi((p-k),(\bar p+k))}{ (p-k)^2})\gamma^{\mu} S(p)
\label{BS0}
\end{equation}
where $S(p)$ and $S(\bar p)$ are the quark propagators, $\sigma$ 
is the string tension, $g$ is the gluon-fermion 
coupling constant and $N$ is the number of colours. In the above we
have also included an additional $g^2 N$ term for the following
reason. It is evident in the theory that the leading UV behaviour is
governed by  $g^2$. Hence to discuss the bound state UV behaviour
we need to consider this contribution too. With our ansatz of $g^2 N$ 
small, we only include the leading UV behaviour  (There are
additional $g^2 N$ terms interfering with $\sigma$ exchange but these
are subleading in the UV regime). In this work we do not consider a 
running $g$  or $\sigma$.

The BS amplitude is decomposed in terms of $4 \times 4$  Dirac matrices \cite{naka,seto} 
\begin{eqnarray}
\phi= \phi_{S} + \gamma_{5} \phi_{P}+\gamma_{\mu} \phi^{\mu}_{V} + \gamma_{\mu} 
\gamma_{5} \phi^{\mu}_{A} + \sigma_{\mu \nu} \phi^{\mu \nu}_{T}
\label{BS1}
\end{eqnarray}
  
Substituting eq.(\ref{BS1}) in the BS equation , we get the following decomposition for the 
scalar, pseudoscalar, vector and pseudovector amplitudes with the propagators given by 
$ S(p)= i (\not p A(p^2) + B(p^2))$, 
$S(\bar p)= i(\not \bar p \bar A(\bar p^2) + \bar B(\bar p^2))$
\begin{eqnarray}
\phi_{S} =4(\bar A A\bar{p}\cdot p + \bar B B) \int \phi_{S} - 2  (\bar A B \bar p_{\mu} + A \bar B  p_{\mu})\int 
\phi^{\mu}_{V} 
\end{eqnarray}

\begin{eqnarray}
\gamma_{\mu} \phi^{\mu}_{V} &=& 4 ( A \bar B \not p + \bar A B \not\bar p) \int \phi_{S} +2 i \bar A A
\epsilon^{  \alpha \mu \beta \delta} \bar p_{\alpha} p_{\beta} \gamma_{\delta} \int \phi_{A \mu} \\ 
\nonumber 
&-& 2 (\bar A A(\bar p_{\mu} \not p - \bar p \cdot p \gamma_{\mu} + \bar \not p p_{\mu}) +\bar B B \gamma_{\mu}) \int \phi^{\mu}_{V} 
\end{eqnarray}

\begin{eqnarray}
\gamma_{5} \phi_{P} &=& 4(\bar A A \bar{p}\cdot p - \bar B B)\gamma_{5} \int \phi_{P} - 2 (\bar A B \bar p_{\mu} - A 
\bar B  p_{\mu}) \gamma_{5} \int \phi^{\mu}_{A} 
\end{eqnarray}

\begin{eqnarray}
\gamma_{\mu} \gamma_{5} \phi^{\mu}_{A} &=& 4 (\bar A B \not\bar p -A \bar B  \not p ) \gamma_{5} 
\int \phi_{P} + 2 i \bar A A
\epsilon^{ \alpha \mu \beta \delta} \bar p_{\alpha} p_{\beta} \gamma_{\delta} \gamma_{5} \int \phi_{V\mu}
\\ \nonumber &-& 2 (\bar A A(\bar p_{\mu} \not p - \bar p\cdot p \gamma_{\mu} + \bar \not p p_{\mu}) - \bar B B 
\gamma_{\mu}) \gamma_{5} \int \phi^{\mu}_{A}
\end{eqnarray} 

The symbol $\int$ stands for the 4-d momentum integral corresponding to the sum of the confining and the gluon 
interactions.
 
\begin{equation}
\int \phi = \int \frac{i ~d^4 k}{ (2\pi)^4}(
 \frac{\sigma N  \phi((p-k),(\bar p+k))}{ (p-k)^4}-
\frac{ g^2 N \phi((p-k),(\bar p+k))}{ (p-k)^2})
 \end{equation}
In addition the tensor components are totally determined by the above components.

Since the momentum of the bound states  has to be time like, we can go  to the centre of mass 
frame of the bound state wherein the total  momentum vector is given by
$q=(q_{0}=M,0,0,0)$ where $M$ is the mass of the bound state. The little group is $SO(3)$. In this frame the 
angular momentum decomposition of the BS amplitude  can be done in 
terms of 3-d solid harmonics and $O(3)$ scalar functions of $p^0,\vec{p}$ in the following manner,

\begin{equation} \phi_{i}^{(j)} = \Phi_{i}^{(j)}(p^0,\vec{p}) y_{jm} 
\label{rep}
\end{equation}
for $i= S,P,0V,0A$ $i.e.,$ Scalar ,Pseudoscalar, time component of Vector
and Pseudovector respectively. The remaining $3-d $ vector components are
decomposed as \cite{lock}

\begin{eqnarray} 
\vec{\phi}^{(j)}_{V}&=& (\vec{p}\Phi_{1V}^{(j)} +\vec{J} \Phi_{2V}^{(j)} + i(\vec{p} \times \vec{J})\Phi_{3V}^{(j)} ) 
y_{jm} \\ \nonumber
&=& \Sigma_{jm\delta } \Phi^{jm(\delta)}_{V} \vec Y^{(j+\delta,1)jm}(\vec p)
\end{eqnarray}

\begin{eqnarray}
\vec{\phi}^{(j)}_{A}&=& (\vec{p}\Phi_{1A}^{(j)}\vec{p} +\vec{J} \Phi_{2A}^{(j)} + i(\vec{p} \times \vec{J})\Phi_{3A}^{(j)}) 
y_{jm} \\ \nonumber
\label{repvec}
&=& \Sigma_{jm\delta } \Phi^{jm(\delta)}_{A} \vec Y^{(j+\delta,1)jm}(\vec p)
\end{eqnarray}

where $\delta=0,1$, $\phi^{\mu}_{V}=(\phi_{0V},\vec{\phi}_{V})$ , $\phi^{\mu}_{A}=(\phi_{0A},\vec{\phi_{A}})$, 
$y_{jm}= 
|\vec{p}|^j Y_{jm}$,where $y_{jm}$ are the solid harmonics, $\vec{J}= \vec{p} \times 
\vec{\nabla_{\vec{p}}}$ 
and $\vec Y^{(j+\delta,1)jm}(\vec p)$ are the 3-d vector spherical
harmonics(\cite{lock}).

The relation between $\Phi_{iV}^{(j)}$ and  $\Phi^{jm(\delta)}_{V}$ is given by \cite{lock}
\begin{eqnarray}
\Phi_{1V}^{(j)}&=&-\sqrt{\frac{j+1}{2j+1}} \Phi^{jm(1)}_{V} +\sqrt{\frac{j}{2j+1}} 
\Phi^{jm(-1)}_{V}\\ 
\Phi_{2V}^{(j)}&=&\sqrt{\frac{1}{j(j+1)}} \Phi^{jm(0)}_{V}\\
\Phi_{3V}^{(j)}&=&-\sqrt{\frac{1}{(j+1)(2j+1)}} \Phi^{jm(1)}_{V} -\frac{1}{\vec p^2} 
\sqrt{\frac{1}{j(2j+1)}}\Phi^{jm(-1)}_{V}\\ \nonumber
\end{eqnarray}

similar equations apply for the Pseudovector part.

\section{Representation of the Wave functions}
In eq.(\ref{rep})-eq.(\ref{repvec}) we have introduced functions of $p^0$ and
$\vec{p}$. They  are Lorentz scalars as they are defined in the rest frame of the bound state. 
A convenient representation is
required to make   our analysis transparent. Consider a
scalar 3-point function. In general this is a scalar function of
momenta associated with three independent Lorentz scalar quantities, namely $p^2$,$\bar p^2$,$q^2$ with
$p_{\mu}-\bar p_{\mu}=q_{\mu}$ owing to momentum conservation. Any
scalar function associated with the 3-point function is a function of
these three variables. There exists a spectral representation for such
a function, that of Deser,$et.al,$ \cite{deser}. In the BS
 wave function we are in a similar situation with one of the
scalar variables namely $q^2$ fixed due to an  eigenvalue
condition. There are many equivalent ways of representing such a spectral
representation. We find the most convenient one is due to \cite{ida},

\begin{eqnarray} 
\phi(\bar{p}^2,p^2) &=& \int_{0}^{1} dy \int_{\delta^2}^{\infty} d\alpha  
\frac{\tilde{\phi}(\alpha,y)} {(p- yq)^2 -\alpha + i \epsilon}
\label{spec} \\
&=&\int_{0}^{1}dy \tilde\Phi((p-yq)^2,y)
\nonumber
\end{eqnarray}
Note that $(p-yq)^2= (1-y)p^2+y\bar{p}^2-y(1-y)q^2$. 

  The spectral function $\tilde{\phi}(\alpha,y)$ in 
general is complex and range of $\alpha$
can be from zero to infinity. For a stable bound state we know from
physical considerations that it has a finite size and for certain
range of energies of the constituents this size is not infinity. The
size of the bound state is defined by the onset of exponential fall off
in co-ordinate space. This is  possible only if the $\alpha$ integration
range is above some positive nonvanishing  quantity $\delta^2$ where $\delta$ is the
inverse of the size of the bound state. In general  $\delta$ many depend on $y$.
Here we will take it to be the minimum possible value in the range of
integration. In WC model\cite{cut} the BS wavefunction can be cast into the above form where
$\delta^{2}(y)$ is fixed in terms of masses of the constituents and $\tilde{\phi}$ is a series in derivatives of
$\delta(\alpha-\delta^2)$. This is also a simple  case of the so called Perturbation Theory
Integral representation \cite{ptir}.
 Substituting  this representation for the each of the
scalar functions eq.(\ref{rep}), we can do the loop momentum integrals by introducing
the appropriate Feynman parameter integrals as shown in detail in \cite{sd}. It is instructive to note
the following self-reproducing property of the solid harmonics
which follows from the defining property \cite{cut}, namely, $\nabla_{\vec{p}}^2
y_{jm}(\vec{p})=0$

\begin{equation}
\int d^3 k F(\vec{k}^2)y_{jm}(\vec{k}+\vec{p})= y_{jm}(\vec{p}) \int d^3 k  F(\vec{k}^2) 
\end{equation}
where $F(\vec{k}^2)$ is a sufficiently well behaved function.

{\bf [S][V] SECTOR:}

\begin{eqnarray} 
\int \tilde{\Phi}_{S}^{(j)} &=& 4(\bar A A \bar p \cdot p +\bar B B) \int \int \tilde{\Phi}_{S}^{(j)}
-2(\bar A B \bar p^{0} + A \bar B p^{0})\int \int \tilde{\Phi}_{0V}^{(j)}
\label{first} \\
 &+& 2 (\bar A B + A \bar B)\vec{p}^2 \int \int \tilde{\Phi}_{1V}^{(j)} \nonumber\\
\int \tilde{\Phi}_{0V}^{(j)} &=& 4(\bar A B \bar p^{0} + A \bar
Bp^{0})\int \int \tilde{ \Phi}_{S}^{(j)} +2 (\bar A A \bar p \cdot p -\bar B B - \bar A A \bar p^{0} p^{0})\int \int
\tilde{\Phi}_{0V}^{(j)} \\ 
\nonumber 
&+& 2 \bar A A (\bar p^{0}+p^{0})\vec{p}^2 \int \int \tilde{\Phi}_{1V}^{(j)}\\ 
\int\tilde{\Phi}_{1V}^{(j)} &=& 4 (\bar A B + A \bar B)\int \int \tilde{\Phi}_{S}^{(j)} -2 \bar A A (\bar 
p^{0}+p^{0})\int \int \tilde{\Phi_{0V}^{(j)}} \\ 
\nonumber 
&+& 2 (\bar A A \bar p \cdot p -\bar B B +2 \bar A A \vec{p}^2) \int \int
\tilde{\Phi}_{1V}^{(j)}\\ 
\nonumber
\end{eqnarray}

{\bf [P][A]-SECTOR:}

\begin{eqnarray}
\int \tilde{\Phi}_{P}^{(j)} & = & 4(\bar A A \bar p \cdot p -\bar B B) \int \int
\tilde{\Phi}_{P}^{(j)} -2(\bar A B \bar p^{0} - A \bar B p^{0}) \int \int \tilde{\Phi}_{0A}^{(j)} \\ 
\nonumber 
&+& 2 (\bar A B -  A \bar B) \vec{p}^2 \int \int \tilde{\Phi_{1A}^{(j)}}\\
\int \tilde{\Phi}_{0A}^{(j)} &=&  4(\bar A B \bar p^{0} - A \bar B p^{0}) \int \int \tilde{\Phi}_{P}^{(j)} +2 (\bar A 
A \bar p \cdot p +\bar B B -  2 \bar A A \bar p^{0}  p^{0})\int \int \tilde{\Phi}_{0A}^{(j)}\\ 
\nonumber 
&+& 2 \bar A A (\bar p^{0}+p^{0})\vec{p}^2 \int \int \tilde{\Phi}_{1A}^{(j)}\\
\int \tilde{\Phi}_{1A}^{(j)} &=& 4 (\bar A B -  A \bar B) \int \int \tilde{\Phi}_{P}^{(j)}  -2 \bar A A (\bar 
p^{0}+p^{0})\int \int \tilde{\Phi}_{0A}^{(j)}\\ 
\nonumber 
&+& 2 (\bar A A \bar p \cdot p +\bar B B +2 \bar A A \vec{p}^2) \int \int \tilde{\Phi}_{1A}^{(j)}\\
\nonumber
\end{eqnarray}

{\bf THE [V][A] Mixed Sector: for j$\geq$1}

\begin{eqnarray}
\int \tilde{\Phi}_{2A}^{(j)}  &=&  2 (\bar A A \bar p \cdot p +\bar B B ) \int \int \tilde{\Phi}_{2A}^{(j)} 
+ 2 \bar A A q^{0} \vec{p}^2 \int \int \tilde{\Phi}_{3V}^{(j)}\\
\int \tilde{\Phi}_{3V}^{(j)}  &=& 2 (\bar A A \bar p \cdot p -\bar B B ) \int \int \tilde{\Phi}_{3V}^{(j)} 
+ 2 \bar A A q^{0} \int \int \tilde{\Phi}_{2A}^{(j)}
\end{eqnarray}

\begin{eqnarray}
\int \tilde{\Phi}_{2V}^{(j)} &=&  2 (\bar A A \bar p \cdot p -\bar B B ) \int \int \tilde{\Phi}_{2V}^{(j)} 
+ 2 \bar A A q^{0} \vec{p}^2 \int \int \tilde{\Phi}_{3A}^{(j)}\\
\int \tilde{\Phi}_{3A}^{(j)} &=& 2 (\bar A A \bar p \cdot p +\bar B B ) \int \int \tilde{\Phi}_{3A}^{(j)} 
+ 2 \bar A A q^{0} \int  \int \tilde{\Phi}_{2V}^{(j)} 
\label{last}
\end{eqnarray}

Where the symbol $\int$ stands for 
\begin{equation}
\nonumber
\int \tilde{\Phi}_{i}^{(j)}= \int_{0}^{1}dy \tilde{\Phi}_{i}^{(j)}((p- yq)^2 ,y))
\end{equation}

and $\int\int$ stands for

\begin{eqnarray}
\int \int \tilde{\Phi}_{i}^{(j)}&=&\frac{\sigma N}{(4\pi)^2} \int_{0}^{1} dx
\int_{0}^{1} \frac{dy}{1-x}( x^{j+1} \tilde{\Phi}_{i}^{(j)}(x(p- yq)^2
,y)-\tilde{\Phi}_{i}^{(j)}((p- yq)^2 ,y))\\
\nonumber
 &+&\frac{ g^2 N}{(4\pi)^2}  \int_{0}^{1} dx \int_{0}^{1} dy \int_{-\infty}^{x(p-yq)^2} d\beta
x^{j} \tilde{\Phi}_{i}^{(j)}(\beta ,y)
\end{eqnarray}
for $i=S,P,0V,0A,2V,2A$. The  $x^{j}$ in the previous equation is replaced by $x^{j+1}$ for $i= 1V,3V,1A,3A$ 
and the equations are written in units of $\frac{\sigma N}{(4\pi)^2}=\bar{\sigma}=1 $. Also we define
$\frac{ g^2 N}{(4\pi)^2} =\bar \alpha$. 

 These coupled integral equations essentially become four different
cases as expected from angular momentum algebra, namely the sum of
two spin $1/2$ and orbital angular momentum $l$, yields total angular momentum $j$ as 
\begin{equation}
j=l\otimes \frac{1}{2}\otimes \frac{1}{2}=l\otimes (0\oplus1)=l \oplus l-1 \oplus l \oplus l+1
\end{equation} 
 This explicit decomposition manifested in eq.(\ref{first})-(\ref{last}) is as far as we know
a new result.

\section{Asymptotic Behaviour}
First we consider the behaviour of the wave function for large space
like $p^2$. Here the wave function is real and probes the short
distance behaviour. The integral equation does not couple the UV
behaviour of the wave function to the IR or intermediate regime of
the theory. The UV behaviour of the wavefunction is determined 
self-consistently by the UV interaction alone. The leading UV behaviour of
$\sigma QCD$
is the same as in
QCD.  Using the asymptotic behaviour of $A(p^2)$ and $B(p^2)$,
\begin{eqnarray}
A(p^2) &\sim& -\frac{1}{-p^2 \sqrt{2 \bar{\alpha} \ln(-p^2)}}
\label{asymA}\\
B(p^2) &\sim&   \frac{ \ln(-p^2)}{-p^2}
\end{eqnarray}
we adopt  the same procedure as shown in \cite{sd}.

The leading order  asymptotic  behaviour of the BS amplitudes in the [P][A] and [S][V] sector are the same. Its in the next to 
leading order(NLO) that they differ.
The leading order BS amplitudes go like,
\begin{eqnarray}
\phi_{P}^{(j)} &\sim &\phi_{S}^{(j)} \sim  \frac{1}{(-p^2)^{j+2} (\ln(-p^2))^{1+\frac{2}{j+1}}}
\label{asymfirst}\\
\phi_{0A}^{(j)} &\sim& \phi_{0V}^{(j)}\sim \frac{1}{(-p^2)^{j+2} (\ln(-p^2))^{1+\frac{1}{j+1}}}\\
\phi_{1A}^{(j)} &\sim& \phi_{1V}^{(j)} \sim \frac{1}{(-p^2)^{j+3} (\ln(-p^2))^{1-\frac{1}{j+2}}}
\end{eqnarray}
 For $j\geq 1$
\begin{eqnarray}
\phi_{2V}^{(j)} &\sim &\phi_{2A}^{(j)} \sim  \frac{1}{(-p^2)^{j+2}(\ln(-p^2))^{1+\frac{1}{j+1}}}\\
\phi_{3V}^{(j)} &\sim& \phi_{3A}^{(j)} \sim \frac{1}{(-p^2)^{j+3} (\ln(-p^2))^{1+\frac{1}{j+2}}}
\label{asymlast}
\end{eqnarray}

 We have not used running $\bar \alpha$ in the above analysis. It is seen
that $A(p^2)$ function dominates as expected and the 
asymptotic  behaviour of the wave function is independent of $\bar \alpha$
due to the dependence of $A(p^2)$ on $\bar \alpha$ as given in
eq.(\ref{asymA}). It is also evident that no further infinite
renormalisations are needed as the BS wavefunction asymptotic behaviour
is sufficiently small that all momentum integrals are finite. This 
 demonstrates that the theory is made
finite by the standard wavefunction, mass and string tension
renormalisations alone. The leading aysmptotic behaviour as shown in  eq.(\ref{asymfirst})-(\ref{asymlast}) is the same
both in $QCD$ and $\sigma QCD$. For 
completeness we mention that if we ignore  the $g^2 N$ term in eq.(\ref{BS0}) 
then the  asymptotic analysis, yields similar results as 
eq.(\ref{asymfirst})-(\ref{asymlast}) with the powers of $p^2$ 
decreased by one and all $\ln(-p^2)$  powers are zero.

\section{Spectrum of Light Mesons}
 The BS equation is simplified considerably in their algebraic complexity. 
Generically it is very much like in the WC model. Major difference 
being that the propagator functions are complicated functions unlike 
simple poles in the WC model. The eigenvalue problem is well defined 
once the explicit $A(p^2)$ and $B(p^2)$ functions of the quark propagator are given.

 The most important properties of $A(p^2)$ and $B(p^2)$ that we exploit 
 is that they are analytic functions near 
$p^2=0$ and the onset of non-analyticity is near the threshold mass
$\tilde m$, $i.e.,$ $p^2 = \tilde{m}^2$ 
in units of $\bar \sigma$. Considering the BS wavefunctions 
(\ref{first})-(\ref{last}) we first note that these functions are 
real and  analytic for $ p^2< \tilde{m}^2 $ and $ \bar{p}^2 < \tilde{\bar m}^2 $.  This is equivalent to saying that 
for all eigenvalues $ q $ such that $ q_{0} < \tilde{m} +\tilde{\bar{m}}$
the wave functions are real and analytic. It is also 
evident from the standard arguments \cite{wick,cut,ida} that for $ q_{0} > \tilde{m} +\tilde{\bar{m}}$  
the wavefunctions are necessarily complex and perhaps even unstable, $i.e.,$ 
the eigenvalue $q_0$ itself may be complex.

 For light quarks where the renormalised mass(current mass) $m$ is much smaller than $\bar \sigma$, we have shown \cite{sd} that
threshold mass(constituent mas) $\tilde m^2 \ll \bar \sigma$, indeed we estimated that $\tilde m^2 \approx .02 \bar \sigma$. 
For stable mesons $q_0< \tilde m +  \tilde{\bar m}$, consequently $q_0 \ll \sqrt{\bar \sigma}$. Therefore all 
stable bound states in this system are necessarily deeply bound.
For simplicity we ignore the 
gluon coupling and keep only the string tension contribution to the BS equation.  We  solve the  BS 
equation at $p^2=\bar{p}^2=0$ where we know explicitly the propagator 
functions. Consider the case when  the BS wavefunction is non 
vanishing at $p^2=\bar{p}^2=0$, since $q^2 \ll 1$, we negelect the $ x q^2$ 
dependence in the  $r.h.s$ of eq.(\ref{first})-(\ref{last}). Then  all $x$ 
integrations can be done explicitly (The $ y$ integration can be done 
formally on both sides).  Consequently the BS equation  reduces to an 
ordinary  matrix eigenvalue equation in each of the different sectors 
at  $p^2=\bar{p}^2=0$. Solution to these homogenous equations is guaranteed 
if the corresponding determinant of the matrix is zero. Noting that all our 
calculations are valid only if $ q^2 > 0 $   the relevant solutions resulting
from the vanishing of the determinant are given below. The explicit answers
are given for renormalised quark mass, $ m = \bar{m}$.
 
% {\bf 1. [S][V] SECTOR:}
\begin{equation}
\mbox{[S][V]~ SECTOR}:~~~   q^{2}=\frac{(2 b_{0}^2 H +1)(4 b_{0}^2 L -1)}{2 a_{0}^2 L(1-2 b_{0}^2 H )}
\end{equation}
 For small quark masses $m$ we have for $j=0$ and $1$, $M^2(j^P)$ where $P$ is the intrinsic parity. 
\begin{eqnarray}
M^2(0^{+}) &\approx& \frac{1}{7}+\frac{153}{196} m\\
M^2(1^{-}) &\approx& \frac{5}{138}+\frac{2171}{6348} m
\label{vect}\\
\nonumber
\end{eqnarray}

% {\bf 2. [P][A] SECTOR:}
\begin{equation}
\mbox{[P][A]~ SECTOR}:~~~ q^2=\frac{(2 b_{0}^2 L -1)(4 b_{0}^2 L +1)}{2 a_{0}^2 L(1+2 b_{0}^2 L )}
\end{equation}
For small quark masses $m$ we have for $j=0$
\begin{equation}
M_{\pi}^2 \equiv M^2(0^{-}) \approx \frac{3}{4}m
\label{pcac}\\
\nonumber
\end{equation}

\begin{displaymath}
\mbox{ [V][A] SECTOR}: \left\{ \begin{array}{l}
q^{2} =\frac{(2 b_{0}^{2} H +1)(2 b_{0}^{2} L -1)}{ a_{0}^{2} (L+ H )}\\
q^{2} =\frac{(2 b_{0}^{2} H -1)(2 b_{0}^{2} L +1)}{ a_{0}^{2} (L+ H )}\\
\end{array}\right.
\end{displaymath}
For small quark masses $m$ we have for $j=1$
\begin{eqnarray}
M^2(1^{+}) &\approx& \frac{23}{160}+\frac{121}{160} m\\
M^2(1^{-}) &\approx& \frac{7}{160}+\frac{13}{32} m\\
\nonumber
\end{eqnarray}
where
\begin{eqnarray}
b_0=\frac{m-\sqrt{m^2+16}}{8};a_0=\frac{b_0}{m-b_0}\\
H=\int_0^1 dx \frac{x^{j+2}-1}{1-x};L=\int_0^1 dx
\frac{x^{j+1}-1}{1-x}\\
\nonumber
\end{eqnarray}
Although the angular momentum $j$ can become arbitrarily large, we
find for larger $j$ than what we have considered, $q^2$ becomes
negative, thus negating our initial assumption that $q_0$ is time
like. Hence these are discarded. 

 In addition we can have solutions where 
the BS wavefunctions vanish at $p^2=\bar{p}^2=0$. Since 
they have to be analytic they have to 
vanish as integer powers of $p^2$ or $\bar{p}^2$. Noting 
that in our representation eq.(\ref{spec}), this 
can only be of the form $((p- y q)^2)^n$ and in the limit of  small $q^2$ this also
approximately vanishes. Implementing these kind of wavefunctions we get precisely the
radial excitations. Trivial algebra shows that approximately the eigenvalues $M^2$ are 
functions of $j+n$ only. All these eigenfunctions can be given Taylor expansions in  $p^2$ and
$\bar{p}^2$ just as we did in \cite{sd} for the quark propagator. This is a
double series and convergence properties of this series is technically
more cumbersome to handle and we have postponed it for later study.

Indeed the above conditions are 
only necessary conditions for the existence of the bound
state. Sufficient conditions have  yet to be stated. In addition to
the above there are many complex solutions with real part of 
$q_0 > \tilde{m} + \tilde{\bar{m}} $. While these are acceptable
as eigenvalue conditions, these should be truly taken as unstable
resonances.

 Many of the eigenvalues both real and complex have to satisfy the
finite size criterion. Namely the size of a bound state or the extent
to which the wave function is spread out should be finite  $i.e.,$  $\delta^2$
in eq.(\ref{spec}) should be non-zero.  We are unable to estimate 
this analytically from the BS equation but a heuristic argument to be stated later suggests
that all eigenvalues with $q^0< \tilde{m}+ \tilde{\bar{m}}$, where
$\tilde{m}$ and $\tilde{\bar{m}}$ are the threshold masses, can exist.

\begin{figure}[h]
\begin{center}
\epsfig{file=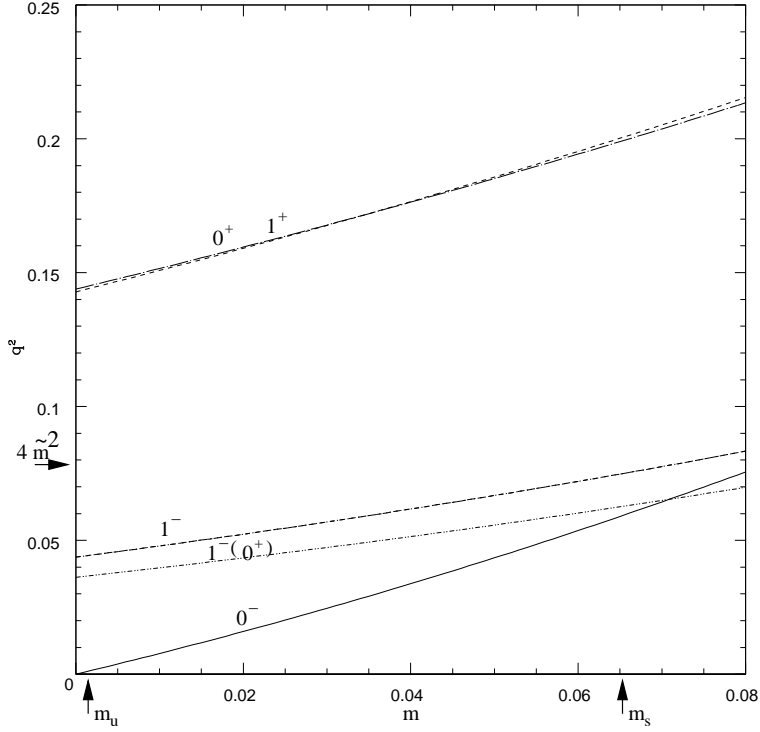,width=10cm,angle=0}
\end{center}
\caption{Meson Spectrum with $j^P$. The one in the bracket is a radial excitation. Only mesons with $q^2< 4 \tilde{m}^2$ are stable}
\label{meson}
\end{figure}

 Let us look at the phenomenological implications of our
spectrum. Fig[\ref{meson}] gives eigenvalue $M^2$ versus the quark mass $m$,
both of which are in units of $\bar\sigma$ and $\sqrt{\bar\sigma}$
respectively. We first compare  pseudoscalar $0^-$ with the rest. This is very well 
understood in QCD \cite{witten}. From first principles
both in continuum and lattice under wide circumstances one can show
that the lightest meson is the pseudoscalar, 
in particular $M(0^-)\leq M(1^-)$. This is also borne out by experimental
data. In our theory this inequality cannot be formally shown to be valid, however it is maintained for renormalised
mass $m$ less than $.07$ and it is disobeyed for larger $m$. So we
have to  conclude from this that this theory is qualitatively  different from QCD for $m>0.07$.

 Next we will take $M_{\pi}$ and $M_{\rho}$ mesons as given to fix physical values for $\bar
\sigma$ and the mass of the up quark, $m_u$(for down quark we take $m_d=m_u$). 
We find from eq.(\ref{vect}) and eq.(\ref{pcac}),  
$\bar\sigma=(4GeV)^2$ and  $m_u=6 MeV$ \cite{leut}. The Scalar $0^+$ turns out to be very
heavy($1.56 GeV$) and cannot certainly be a stable bound state as it
is greater than $2\tilde m_u$, where $\tilde m_u$ is the threshold
mass of the $up$ quark which we estimate in this model to be about
$.56GeV$. This implies all $up$ and $down$ quark bound states less
than $1.2GeV$ are stable.

        In the strange quark sector, the $1^-$ vector bound state of
$s\bar{s}$ is unambigously known to be $\phi(1020)$. From Fig[\ref{meson}], it can
be inferred that the strange quark mass($m_s$) is about $0.26GeV$ and
a pseudoscalar $0^-$ will have a mass $0.95GeV$ which corresponds to
the  $\eta^{'}$ meson. That is, in this scenario most of the mass of the
$\eta^{'}$ meson can be thought of as coming from $s\bar s$(flavour mixing
is not attempted in our calculations). In our model strange quark
mass is very close to the crossover regime where $1^-$ and $0^-$
cross, beyond which the model qualitatively fails to be QCD like. 

 Eq(\ref{pcac}) demonstrates the well known consequence of PCAC 
namely, the square of the mass of the pion is proportional to the 
current quark mass $m$ and the proportionality constant in this 
model is $\frac{3}{4} \sqrt{\bar \sigma}=3 Gev$. Furthermore 
we see that  there are two states way above the threshold, 
namely $0^+$ and $1^+$.  $0^+$ state is the well known 
"$\sigma$ particle". It is clearly extraordinarily massive and is
expected to be unstable even in this lowest order 
calculation as it exceeds the threshold energy.  

  Physical spectrum is expected to show a  lot of mixing
in flavour neutral particles. This can be anticipated in this model
purely because the threshold mass for all the flavours is about the
same \cite{sd}, hence $1 \over N$ corrections can become dominant due to
kinematical reasons alone . By carrying out $1 \over N$ calculation we can have  
better fit to phenomenology.
        
        Now we  make a semi-analytic discussion as to the 
size of the bound states from general considerations of BS
equation. These equations have a generic form as
\begin{equation}
\phi=S \bar{S} \int  K \phi
\end{equation}
where $S(p)$ and  $\bar{S}(\bar{p})$ are the propagator functions of the
constituents and $K$ is the interaction kernel. In general spatial
length scales can be present from the interaction kernel. For our
discussion we shall assume that the interaction kernel has long range
like the massless gluon. Even the string tension term is long ranged.
For these type of kernels, the length scales come from the propagators
of the constituents alone. For light quarks these are complicated and
not so well known functions. however we do know that they have
spectral representations starting from a threshold mass $\tilde{ m}^2$ and
$\tilde{ \bar{m}}^2$. Hence the smallest mass scale entering the equation comes through
 this threshold mass. A crude approximation of the propagators for 
$p^2<\tilde{ m}^2$ and $\bar{p}^2<\tilde{ \bar{m}}^2$ would be, 
\begin{equation}
S(p)~ \bar S(\bar p) \approx  \frac{1}{p^2-\tilde m^2}~ \frac{1}{\bar{p}^2-\tilde
{\bar{m}}^2}
\end{equation}
where $\bar{p}=p-q$. This is qualtitatively reasonable but not quantitatively.
 In the rest frame let us anticipate that there is
an average energy $p^0$ for quark and $\bar{p}^0$ for antiquark. This can be
estimated to be 
\begin{equation}
p_0\approx\frac{\tilde m}{\tilde{m}+ \tilde{\bar m}}q_0 \;\; \mbox{and} \;\;\;
q_0-p_0\approx \frac{\tilde{\bar m}}{\tilde{m}+ \tilde{\bar m}}q_0
\end{equation}
then we have
\begin{equation}
S(p)~ \bar S(\bar p)\approx \frac{1}{(\vec{p}^2 +\frac{\tilde{
m}^2}{(\tilde{m}+ \tilde{\bar m})^2}((\tilde{m}+ \tilde{\bar{m}})^2
-q_{0}^2))}~ \frac{1}{(\vec{p}^2 +\frac{\tilde{\bar{m}}^2}{(\tilde{m}+
\tilde{\bar m})^2}((\tilde{m}+ \tilde{\bar{m}})^2-q_{0}^2))}
\label{length}
\end{equation}

Hence naturally in eq.(\ref{length}) the largest length  scale given by the
exponential fall off of the wave function in position space is set by
\begin{equation} 
\delta^2 = min(\tilde{m}^2, \tilde{\bar{m}}^2) (1-\frac{q_{0}^2}{(\tilde{m}+ \tilde{\bar{m}})^2})
\end{equation}
Where $\frac{1}{\delta}$ is the size of the system. For quarks this is
an estimate since the propagator is not a simple pole. But the
existence of the
spectral representation for  quark propagators seems to indicate that
it is a good estimate. For deeply bound states the size is entirely determined
by the threshold mass. The above estimate holds good exactly for long range interacting non-relativistic
systems such as coulomb interactions. 

 This estimate also suggests that when $q_0$ reaches threshold,
$\delta$ vanishes suggesting that the bound system attains infinite
size. This simple consideration is always valid as an estimate
whenever there are no massive exchange interactions in the interaction
kernel. Indeed when there are such interactions the largest length
scale between the propagators and the interactions (or small mass
$\delta$ scale) dominates in dictating the size of the system.
From the simple minded considerations above we can conclude that 
massless particles cannot be bound as it will necessarily have
infinite size. We make an interesting observation about chiral 
symmetric phase of quarks at zero temperature. In this phase the
quark has vanishing threshold  mass hence from our considerations it cannot be bound.
Chiral symmetric vaccum is automatically a non-confining vaccum.

\section{Discussion}
 We have discussed a complete relativistic description of bound states and 
the BS equation is reduced to a set of simpler equations. The form of the 
equations(\ref{first})-(\ref{last}) is valid for a general chiral symmetric 
interaction kernel. Many of our later algebraic simplifications is due to the
absence of a mass scale in the interaction kernel however our method  can  handle 
even if there is a mass scale in the interaction kernel.

  We have concentrated mostly on tightly bound systems primarily because
the string tension $\bar \sigma$ is large in $\sigma QCD$. Consequently the tight 
binding approximation is relevant. For a range of low quark masses the 
lightest meson is the pseudoscalar which extrapolates all the way to the 
Goldstone mode. We verified the PCAC result that the pion mass $M_\pi$ is 
related to the renormalised quark mass $m$ as shown in eq.(\ref{pcac}).

 On general grounds we find that there are very few stable mesons. This follows 
entirely from the fact that threshold masses for light quarks $\tilde m$ are much 
smaller than the string tension $\bar \sigma$. The BS equations can be studied
to look for complex eigenvalues and thus the unstable mesons. We did find several
complex eigenvalues numerically to the set of equations (\ref{first})-(\ref{last}). We are as 
yet unable to systematically analyse them. Primary reason being that the method 
of finding the spectrum  is necessary but not sufficient.  Another important 
necessary condition we can argue is that of the size of the bound states.
For stable bound states the argument presented earlier  is good enough but for 
unstable bound states this needs to be improved upon.

 Another alternative is to invent a formal series solution as done in
WC model. In principle this method can be applied here as well but the tedium makes the
physics non-transparent. Our analytic method of computing the tightly bound 
spectrum (low lying tightly bound states when the coupling is large) was applied to
WC model and we reproduced the known conventional results \cite{thesis}.

  Fitting  experimental data to this model has shown that the threshold mass for $u,d$ or $s$ 
quarks is about the same because of the string tension $\bar \sigma$ being so large.
Consequently it is easy to anticipate that there can be large flavour mixing.
In our model this is next order in $1\over N$. Consequently  we  expect 
$1\over N$ corrections are not necessarily small for light quarks.

 There are several normalisation schemes\cite{pred} for the BS wave function such as
Cutkosky, charge, energy normalisation conditions. One of the primary
drawbacks here is that all known normalisations for relativistic bound
states are  not positive norms in the standard sense. Consequently
they are not of much utitlity. However it has been shown that all the known
normalisations are equivalent \cite{pred}.

  We have not dealt with heavy quarks for they fall in a different class
altogether. In this model string tension decreases at larger energy scales.
So the value of string tension is indeed much smaller for heavy quarks and
thus they fall into the loosely bound regime. That is the binding energy
is much smaller than the rest mass or the threshold mass. This is precisely
the non-relativistic limit. If we perform the standard non-relativistic
approximation to the equations (\ref{first})-(\ref{last}), we do get the standard Schrodinger
picture \cite{god} in momentum space along with spin-orbit interactions. 

%We however would like to comment that in making this approximation we take the interaction to be instantaneous 
%instead of the causal interaction as studied in this  paper. This is not a well
%understood approximation. Tiktopaulos\cite{tikt} made an interesting observation that with a 
%norm that he proposes the non-relativistic result can be understood as a bound 
%on the spectrum. This idea needs to be developed a little more precisely.

  $\sigma QCD$ model has many features of QCD as we explicitly 
emphasised in our papers \cite{ram,sd}. Yet we have
shown that for renormalised quark mass(current mass) $m>.07\sqrt{\bar\sigma}$ we disobey a well known 
inequality of the meson spectrum which is understood theoretically and valid
experimentally, namely in any flavour sector the pseudoscalar is the lightest meson. 
This follows purely from the positivity properties of QCD in the Euclidean
formulation. Analogous positivity  property is not valid in our model. But it is interesting to 
note that it is  of no consequence for all
light  quarks($u$, $d$ and  $s$). 

 A crude estimate suggests that for heavy quarks in our model $i.e.,$ for
$m>3   \sqrt{\bar\sigma}$ we recover   the pseudoscalar mass inequality. If we consider
that $\bar \sigma$ is small for heavy quarks we do envisage that charm, bottom,
top can also be accomodated. This will be discussed in a later publication. But we 
have to bear in mind that there is a range of quark masses which will disobey the light
pseudoscalar mass
inequality and the quarks that occur in nature are not in that regime.

  We have presented the spectrum calculation explicitly for the case where 
both flavours have the same renormalised mass. We can also do these 
calculations analytically if they are unequal.  For the deeply
bound states that we have considered the effect is small, comparable to
$1 \over N$ corrections. Finally many of our results can be applied in the
context of technicolour scenarios as well.

\end{document}